\documentclass[aps,twocolumn,pra,showpacs,floatfix,superscriptaddress]{revtex4-1}

\usepackage{amssymb}
\usepackage{amsmath}
\usepackage{graphicx}

\newcommand{\tm}{\tablenotemark}
\begin{document}

\title{Testing atomic wave functions in the nuclear vicinity:
the hyperfine structure with empirically-deduced nuclear and quantum electrodynamic effects}
\author{J. S. M. Ginges}
\affiliation{School of Mathematics and Physics, The University of
  Queensland, Brisbane QLD 4072, Australia}
\affiliation{Centre for Engineered Quantum Systems, School of Physics,
The University of Sydney, Sydney NSW 2006, Australia}
\email{j.ginges@uq.edu.au}
\author{A. V. Volotka}
\affiliation{Helmholtz-Institut Jena, Fr\"obelstieg 3, D-07743 Jena,
  Germany}
\date{\today}

\begin{abstract}
Calculations of the magnetic hyperfine structure rely on the input of
nuclear properties --  nuclear magnetic moments and nuclear
magnetization distributions -- as well as 
quantum electrodynamic (QED) radiative corrections for high-accuracy
evaluation in heavy atoms. The uncertainties associated with 
assumed values of these properties limit the accuracy of 
hyperfine calculations. 
For example, 
for the heavy alkali-metal atoms Cs and Fr,
these uncertainties may amount collectively to almost 
1\% or 2\%,
respectively.
In this paper we propose a method for removing the dependence of
hyperfine structure calculations on assumed values of 
nuclear magnetic moments and nuclear magnetization distributions
by determining these effects empirically from measurements of 
the hyperfine structure for high states. 
The method is valid for $s$, $p_{1/2}$, and $p_{3/2}$ states of
alkali-metal atoms and alkali-metal-like ions.
We have shown that for $s$ states the dependence on QED 
effects may also be removed to high accuracy.
The ability to probe the electronic wave functions, through
hyperfine comparisons, with significantly increased accuracy is important for the analysis of
atomic parity violation measurements and may enable the accuracy of atomic parity violation
calculations to be improved. 
More broadly, it opens the way for further development of high-precision atomic many-body
methods.

\end{abstract}


\maketitle

\section{Introduction}

Studies of atomic parity violation provide important low-energy tests of the
standard model of particle physics. The largest effect in atoms arises from the nuclear weak
charge which depends on a unique combination of fundamental coupling
constants. This makes atomic parity violation measurements uniquely
sensitive to certain types of new physics, complementing searches for
new physics performed at high energies \cite{particle1,particle2,particle3}. 

Extraction of the nuclear weak charge from atomic measurements
requires high-precision atomic calculations (see, e.g., reviews 
\cite{review1,review2}). 
To gauge the accuracy of
these calculations, comparison of theoretical and experimental
determinations of electric dipole transition amplitudes, energy
intervals, and hyperfine structure constants is made \cite{dfg2002,pbd2009}. 
The hyperfine structure is sensitive to the atomic wave functions in
the nuclear region, and it is the hyperfine structure for which the 
largest deviations are seen. 

The atoms and ions of interest for atomic parity violation measurements
include Cs \cite{cs_pnc,cs_pnc2}, Fr \cite{fr_pnc1,fr_pnc2,fr_pnc3}, Ba$^{+}$
\cite{ba_pnc1,ba_pnc2}, and Ra$^{+}$ \cite{ra_pnc}. 
The highest precision in atomic parity violation studies has been reached for
$^{133}$Cs, the measurement accurate to 0.35\% \cite{wieman} and calculations
accurate to within 0.5\% \cite{pbd2009,DBFR_pnc,dfg2002,radpot}. 
The future of nuclear spin-independent atomic parity
violation studies on a single isotope depends on the ability to control the error of
atomic calculations to $\approx 0.1\%$.

In the current work, we pave the way for significantly improved
understanding of the electronic wave functions in the nuclear region.
We propose a method for empirically 
correcting the unknown or neglected nuclear properties and quantum electrodynamic (QED)
radiative corrections in 
hyperfine structure calculations by exploiting the scaling of
different effects for higher states. 

Beyond the uncertainties associated with many-electron correlations 
in the atomic theory evaluation of the hyperfine structure, there are 
several other sources of uncertainty related to assumed values of (i) the nuclear magnetic
moment, (ii) the finite-magnetization distribution of the nucleus, and (iii) QED radiative 
corrections (or their neglect). 
The size of the error associated with each of these may be 
several 0.1\%, or even $\approx 1\%$, for the systems of interest for 
parity violation studies. Controlling these errors is
crucial if $0.1\%$ tests of the atomic theory is to be made.

Indeed, (i) in the comparison between theory and experiment for the hyperfine
structure, 
a value for the nuclear magnetic moment is assumed. However, for the 
Fr isotopes, these are not known to better than 1-2\% \cite{fr211,gomez}.
(ii) An assumed magnetization distribution is used in atomic
calculations, the most routinely-used being the uniformly magnetized 
sphere. 
There is data, however, for the hyperfine structure 
for the neutron-deficient isotopes of 
Fr \cite{Fr_anomaly1,Fr_anomaly2} 
that supports the validity of the single-particle model for that system. 
In our recent work on the ground-state hyperfine
structure \cite{GVF_QED_BW}, we demonstrated that using 
the single-particle model 
gives a result for $^{211}$Fr that differs by 1.3\% 
from that found using the sphere; 
for $^{133}$Cs this difference is $0.5\%$. 
(iii) Rigorous calculations of QED radiative corrections to the
ground-state hyperfine structure have been performed at the one-loop level for the alkali-metal atoms 
\cite{SC_hfs,GVF_QED_BW} and only recently for alkali-metal-like ions
Ba$^+$ and Ra$^+$ \cite{GVF_QED_BW}, with contributions entering at
around $0.5\%$ for Cs, Ba$^+$, Fr, Ra$^+$. Overwhelmingly, these
effects have been neglected or crudely estimated in theoretical 
determinations of the hyperfine structure. 

The uncertainty associated with the nuclear magnetization distribution also poses a 
problem in the area of QED tests in few-electron highly-charged
ions. That problem is addressed by constructing a difference between
hyperfine intervals of the ion in question and the hydrogen-like ion
that cancels the effect \cite{shabaev_diff}.
This difference, however, relies on precise knowledge of the nuclear
magnetic moment. This proved problematic in studies of Li-like
$^{208}$Bi, where the accepted value of this moment turned
out to be inaccurate \cite{skripnikov}.

\section{The hyperfine structure across principal quantum number}

\subsection{Factorizing the hyperfine structure}

The magnetic hyperfine interaction is given by
\begin{equation}
\label{eq:hfs}
h_{\rm hfs}=c{\boldsymbol \alpha}\cdot {\bf A}
=\frac{1}{c}\frac{{\boldsymbol \mu}\cdot ({\boldsymbol r}\times
{\boldsymbol \alpha})}{r^3}F(r) \ ,
\end{equation}
where ${\boldsymbol \alpha}$ is a Dirac matrix, 
${\bf A}$ is the nuclear vector potential,
${\boldsymbol \mu}=\mu {\bf I}/I$ is the nuclear magnetic moment 
and ${\bf I}$ is the nuclear spin,
and $F(r)$ describes the magnetization distribution. 
We use atomic units ($|e|=m=\hbar=4\pi \epsilon_0=1$, $c=1/\alpha$).
$F(r)=1$ corresponds to the
case of point-nucleus magnetization. 
For a finite-nucleus magnetization distribution, the value $F(r)-1$ differs 
from zero within the nucleus, $r\le
r_n$. Account of finite-nucleus magnetization gives a correction to
the (point-magnetization) hyperfine structure -- the
Bohr-Weisskopf (BW) effect \cite{BW}.   

The splitting due to the magnetic hyperfine interaction (\ref{eq:hfs})
may be quantified in terms of the
magnetic constant $A$. In the zeroth-order approximation
(lowest-order in the atomic potential and for point-nucleus 
magnetization), the hyperfine $A$ constant for the state with principal quantum
number $n$ and angular momentum quantum number $\kappa$ ($\kappa =
-1, 1, -2,...$ for $s$, $p_{1/2}$, $p_{3/2}$,...) is given by
\begin{equation}
A_{n\kappa} = -\frac{\alpha^2}{m_p}\frac{g_I\kappa}{J(J+1)} \int_0^{\infty}dr\,f(r)g(r)/r^2 \ .
\end{equation}
Here $J$ is the electronic angular momentum, 
$m_p$ is the proton mass, $g_I=\mu/(\mu_N I)$ is the nuclear
g-factor, and 
$f(r)$ and $\alpha g(r)$ are the upper and lower radial components
of the relativistic wave functions $\varphi$  
that satisfy the Dirac equation
\begin{equation}
\label{eq:dirac}
\big(c{\boldsymbol \alpha}\cdot {\bf p}+(\beta-1)c^2 + V_{\rm
  nuc}(r)+V_{\rm el}\big) \varphi = \epsilon \varphi \ ,
\end{equation}
where $\beta$ is a Dirac matrix. 
In our many-body calculations for the hyperfine structure, we use 
the Hartree-Fock potential as our starting potential, $V_{\rm
  el}=V_{\rm HF}$. 
Finite nuclear charge distribution is included in the determination of the wave
functions, with $V_{\rm nuc}$
corresponding to a 2-parameter Fermi distribution in our calculations.

We introduce the following parameterization for the hyperfine $A$ constant,
\begin{equation}
\label{eq:scaling_lower}
A_{n\kappa}=A_{n\kappa}^{\rm MB}\frac{\mu}{\mu_N}\Big(1+\frac{\alpha}{\pi} F^{\rm
  BW}_{n\kappa}+\frac{\alpha}{\pi} F^{\rm QED}_{n\kappa}\Big) \ .
\end{equation}
We explicitly separate different aspects of the hyperfine problem. The
nuclear magnetic moment $\mu$ is factored out and the value for 
$A_{n\kappa}^{\rm MB}$ comes only from electronic many-body calculations
with the nuclear magnetic moment set to $\mu_N$ and for point-nucleus 
magnetization.      
The parameters $F_{n\kappa}^{\rm BW}$ and $F_{n\kappa}^{\rm QED}$ give the relative
Bohr-Weisskopf correction and the relative QED correction, respectively.
In the following, we will consider how the relative correlation, BW, 
and QED corrections scale with principal quantum number.

\subsection{Correlation corrections}

Our many-body calculations are carried out using the correlation potential 
approach \cite{DFSS1987}. A non-local, energy-dependent correlation potential
$\Sigma({\bf r},{\bf r}',\epsilon)$ 
is constructed such that, in lowest order, the average value of this potential 
coincides with the second-order correlation correction to the energy. 
We use the Feynman diagram technique to include into $\Sigma$ the electron-electron
screening and the hole-particle interaction to all orders in the
Coulomb interaction \cite{DFSS1988}. This potential is added to the relativistic 
Hartree-Fock equation (\ref{eq:dirac}), with $V_{\rm HF}\rightarrow V_{\rm
  HF}+\Sigma^{(\infty)}$, and correlation-corrected (Brueckner) energies $\epsilon_{\rm Br}$ 
and orbitals $\varphi_{\rm Br}$ are obtained.

The dominant part of the external-field correlation corrections -- the core polarization -- is included using 
the random-phase approximation with exchange (RPA). From this we get a
correction to the hyperfine operator which corresponds to a
hyperfine-modified Hartree-Fock potential, $h_{\rm hfs}+\delta V_{\rm
  hfs}$ \cite{DFSS1987}. Inclusion of the correlation potential and RPA corrections 
corresponds to evaluation of the matrix element $\langle \varphi_{\rm Br}|h_{\rm
  hfs}+\delta V_{\rm hfs}| \varphi_{\rm Br}\rangle$.

\begin{figure}[bpt]
\begin{center}
\includegraphics[width=\columnwidth]{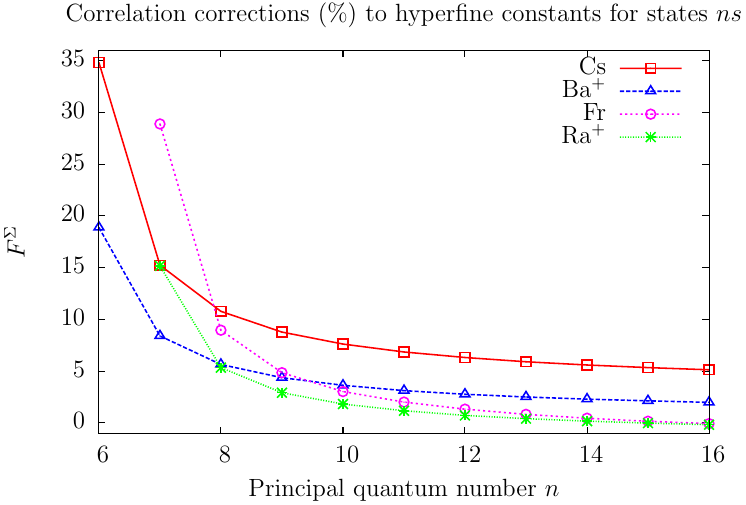}
\caption{Relative correlation corrections $F^{\Sigma}$ to the
  hyperfine structure (in $\%$) for Cs, Ba$^+$, Fr, Ra$^+$. 
This corresponds to evaluation of $\langle \varphi_{\rm Br}|h_{\rm hfs}+\delta V_{\rm hfs}| \varphi_{\rm
  Br}\rangle - \langle \varphi|h_{\rm
  hfs}+\delta V_{\rm hfs}| \varphi\rangle$ relative to 
$\langle \varphi|h_{\rm hfs}+\delta V_{\rm hfs}| \varphi\rangle$.}
\label{fig:corr}
\end{center}
\end{figure}

In Fig.~\ref{fig:corr} we plot the relative correlation corrections for
$ns$ states from the ground state to principal quantum number $n=16$
for alkali-metal atoms and alkali-metal-like ions of
interest for parity violation studies, Cs, Fr, Ba$^+$, Ra$^+$.
These corrections correspond to the difference 
$\langle \varphi_{\rm Br}|h_{\rm hfs}+\delta V_{\rm hfs}| \varphi_{\rm
  Br}\rangle - \langle \varphi|h_{\rm
  hfs}+\delta V_{\rm hfs}| \varphi\rangle$ relative to 
$\langle \varphi|h_{\rm hfs}+\delta V_{\rm hfs}| \varphi\rangle$, which
we denote by $F^{\Sigma}$.
Since most of the uncertainty in many-body calculations is associated 
with evaluation of correlations, the smaller the relative size of
the correlations $F^{\Sigma}$, the smaller the anticipated error in the many-body
calculations. 
Our calculations for smaller correlation corrections (structural
radiation and normalization of states, that enter at around 1\% or
less and are not included in $\langle \varphi_{\rm Br}|h_{\rm
  hfs}+\delta V_{\rm hfs}| \varphi_{\rm Br}\rangle$) also demonstrate a reduction in
relative size with increase in $n$. 
We may therefore expect that significantly higher accuracy may be achieved
for calculations of $A^{\rm  MB}_{n'\kappa}$ for the higher states
compared to calculations of $A^{\rm
  MB}_{n\kappa}$ for the ground or lower level, where $n'>n$ and particularly for $n'\gg n$. 

\subsection{Bohr-Weisskopf corrections}

We have studied the Bohr-Weisskopf effect in two very different models:
(i) the uniform spherical 
distribution, where $F(r)= (r/r_n)^3$, and (ii) the nuclear
single-particle model, with spin nucleon $g$-factors found from
measured nuclear magnetic moments. We refer the reader to, e.g.,
Ref.~\cite{volotka} for the 
single-particle model expressions for $F(r)$, derived in Refs.~\cite{BW,lebellac,shabaev,STKAY}. 
We performed calculations for $^{133}$Cs, $^{211}$Fr, $^{135}$Ba$^+$,
and $^{225}$Ra$^+$ with $2\times10^4$ grid points and a grid radius of
$r=230$\,a.u. for the atoms and $r=120$\,a.u. for the ions. 
Calculations were performed for states $ns$, $np_{1/2}$, and $np_{3/2}$, 
with $n=6-10$ for Cs and Ba$^+$ and $n=7-11$ for Fr and Ra$^+$. Account of
core polarization is important for all non-$s$ states, and for 
states with $j>1/2$ it is responsible for the effect entirely.
We observe that for the $s$, $p_{1/2}$, and $p_{3/2}$ states the
Bohr-Weisskopf effect $(\alpha/\pi)F^{\rm BW}_{n\kappa}$ stays the
same across $n$ to within 0.05\% for all considered
isotopes and magnetization models (the variability tends to be
significantly smaller for the $s$ and $p_{1/2}$ states). 
For example, in the nuclear single-particle model for $^{133}$Cs, the 
values for $F^{\rm BW}$ at the RPA level are $-0.898$, $-0.056$, $-0.247$ for $6s$,
$6p_{1/2}$, $6p_{3/2}$ states and the variation across $n=6-10$ for 
$\alpha/\pi\, F^{\rm BW}$ is $\ll0.1\%$; in the
spherical model the values are about 3.4 times larger. 
Account of correlations hardly influences the $F^{\rm BW}$ values for $s$ states
and changes the values for $p$ states only on the level of $1\%$,
while the variation across $n$ stays about the same. 
Note that in Rb the $n$-independence of the Bohr-Weisskopf 
effect for $s$-states has been observed experimentally across $n=5-7$ 
through isotopic ratios of the hyperfine constants \cite{Rbanomaly}.

Therefore, for the excited states $n's_{1/2}$, $n'p_{1/2}$, and $n'p_{3/2}$  
we may write the hyperfine constant as
\begin{equation}
\label{eq:scaling_higher}
A_{n'\kappa}=A_{n'\kappa}^{\rm  MB}\frac{\mu}{\mu_N}\Big(1+\frac{\alpha}{\pi}F_{n\kappa}^{\rm BW}+
\frac{\alpha}{\pi}F_{n'\kappa}^{\rm QED}\Big) \ ,
\end{equation}
where the relative BW effect coincides with
that for the ground state.
Combining Eqs.~(\ref{eq:scaling_lower}),\,(\ref{eq:scaling_higher}),
it is seen that {\it the dependence on nuclear properties may be
  completely eliminated}, 
\begin{equation}
\label{eq:ratio_gen}
A_{n\kappa}= A^{\rm MB}_{n\kappa} \Big( A_{n'\kappa}^{\rm exp}/A^{\rm MB}_{n'\kappa}\Big)
\Big( 1-\frac{\alpha}{\pi}\delta F^{\rm QED}_{\kappa}\Big) \ .
\end{equation}
We have replaced the total hyperfine constant $A_{n'\kappa} $ with its
experimentally-determined value $A_{n'\kappa}^{\rm exp}$ and 
$\delta F^{\rm QED}_{\kappa}=F^{\rm QED}_{n'\kappa}-F^{\rm
  QED}_{n\kappa}$. The nuclear physics input 
which has until now been required in hyperfine structure calculations may 
instead be determined implicitly through hyperfine structure measurements
for high states, electronic many-body calculations for high states, and the difference in QED 
contributions for the states being compared. We are confident that these may all be
determined with sub-0.1\% uncertainty.

\subsection{QED radiative corrections}

We have gone further in this work and explicitly calculated the QED radiative
corrections for $ns$ states for $^{133}$Cs
over $n=6-10$. Rigorous calculations of the one-loop self-energy and 
vacuum polarization corrections were performed using the extended Furry 
picture with the core-Hartree (CH) potential and the Kohn-Sham (KS)
potential generated for the ground state, as in
Ref.~\cite{GVF_QED_BW}. 
In Table~\ref{tab:QED} we present our results. The
correction changes relatively significantly from 
$n=6$ through to $n=10$ (20\% for CH and 17\% for KS), however a difference of 20\%
in $F^{\rm QED}$ corresponds to $|(\alpha/\pi)\delta F^{\rm QED}_{-1}|\approx 0.05\%$, 
and we expect similar QED differences across $n$ for $s$ states of other heavy alkali-metal atoms and
alkali-metal-like ions.

\begin{table}[bth]
\caption{Relative QED radiative corrections $F^{\rm QED}$ to 
hyperfine constants for $ns$ states of $^{\rm 133}$Cs 
found in core-Hartree and Kohn-Sham potentials, $V_{\rm CH}$ and
$V_{\rm KS}$.} 
\label{tab:QED}
\begin{ruledtabular}
\begin{tabular}{lccccc}
& \multicolumn{5}{c}{$F^{\rm QED}$}\\
 & $6s$ & $7s$ & $8s$ & $9s$ & $10s$ \\
 \hline
$V_{\rm CH}$ & -1.64 & -1.49 & -1.44 & -1.39 & -1.31\\
$V_{\rm KS}$ & -1.91 & -1.77&-1.71 &-1.65 & -1.59 \\
\end{tabular}
\end{ruledtabular}
\end{table}

\subsection{The hyperfine ratio method}

Therefore, we have shown -- accurate to within 0.1\% -- that we
may simplify Eq.~(\ref{eq:ratio_gen}) further for the $s$ states,
\begin{equation}
\label{eq:ratio}
A_{ns}=A_{ns}^{\rm MB}\big(A_{n's}^{\rm exp}/A_{n's}^{\rm
  MB}\big) \ .
\end{equation}
(We replace the notation $A_{n\,-1}$ with $A_{ns}$.) 
The ratio in brackets in Eq.~(\ref{eq:ratio}) corresponds to the nuclear and QED aspects of
the problem, $\mu/{\mu_N}(1+\alpha/\pi\, F^{\rm BW} + \alpha/\pi\,
F^{\rm QED} )$.

\begin{table}[bth]
\caption{Magnetic hyperfine A constants for $6s$ and $7s$ states of 
  $^{133}$Cs from Refs.~\cite{dfg2002,pbd2009}. 
Raw values and percentage deviations (\% dev.) from experiment 
  are shown in the first rows of data, values with QED and
  Bohr-Weisskopf (single-particle Woods-Saxon) corrections from Ref.~\cite{GVF_QED_BW}
  are shown in the following rows, 
and values for $6s$ using the ratio involving $7s$, $(A^{\rm
  exp}_{7s}/A^{\rm th}_{7s})\,A^{\rm th}_{6s}$, and vice-versa for $7s$, are shown in the final rows.  
Units: MHz.}
\label{tab:ratio}
\begin{ruledtabular}
\begin{tabular}{lcccc}
Reference & $A_{6s}$ & \% dev. & $A_{7s}$ & \% dev. \\ 
\hline
Raw & & & & \\
Theory~\cite{dfg2002}, unfitted &2315.0&0.73&545.3&-0.09 \\
Theory~\cite{dfg2002}, fitted &2300.3&0.09&543.8&-0.37\\
Theory~\cite{pbd2009} &2306.6&0.37&544.59&-0.23\\
With QED, BW(SP) & & & & \\
Theory~\cite{dfg2002}, unfitted &2318.4&0.88&546.1&0.06 \\
Theory~\cite{dfg2002}, fitted &2303.8&0.24&544.6&-0.22\\
Theory~\cite{pbd2009} &2319.7&0.94&547.69&0.34\\
Ratio & & & & \\
Theory~\cite{dfg2002}, unfitted &2317.1&0.82&541.3&-0.82\\
Theory~\cite{dfg2002}, fitted &2308.8&0.46&543.3&-0.46\\
Theory~\cite{pbd2009} &2311.8&0.59&542.61&-0.59\\
Experiment &$2298.157$\tm[1]&&$545.818(16)$\tm[2] & \\
\end{tabular}
\end{ruledtabular}
\tablenotetext[1]{Reference~\cite{arimondo}.}
\tablenotetext[2]{Reference~\cite{yang}.}
\end{table}

The right-hand-side of Eq.~(\ref{eq:ratio}) contains the ratio of 
the results of two many-body calculations, $A_{ns}^{\rm
  MB}/A_{n's}^{\rm MB}$. We may write
\begin{equation}
A_{ns}^{\rm MB}\approx A_{ns}^{\rm
  HF}\big(1+F_{ns}^{\delta V}\big)\big(1+F_{ns}^{\Sigma}\big) \ ,
\end{equation}
where $A_{ns}^{\rm HF}$ is the hyperfine constant found in the 
relativistic Hartree-Fock approximation and  $F_{ns}^{\delta V}$, $F_{ns}^{\Sigma}$ 
are the relative RPA and correlation corrections.
The RPA correction is essentially the same for all principal
quantum numbers -- for Cs, $F_{ns}^{\delta V}\approx 0.2$ -- and it 
cancels in the ratio Eq.~(\ref{eq:ratio}). 
It means that the ratio method is largely 
insensitive to account of core polarization.

The atomic theory error, however, is mainly associated with the evaluation of the  
electron-electron correlations, most of which may be represented by  
a correlation potential $\Sigma$. For the correlation
potential, and smaller  
correlation corrections such as the structural radiation and normalization of states, 
the relative correction is not the same for different $n$ -- as we saw
earlier -- 
and we have 
\begin{equation}
\label{eq:diffsig}
A_{ns}^{\rm MB}/A_{n's}^{\rm MB}\approx A_{ns}^{\rm HF}/A_{n's}^{\rm HF}\, 
\big(1+F_{ns}^{\Sigma} - F_{n's}^{\Sigma}\big) \ .
\end{equation}
The ratio therefore depends on the difference in the relative
correlation corrections between states $ns$ and $n's$, 
$F_{ns}^{\Sigma} - F_{n's}^{\Sigma}$. If we are 
interested in preserving the dependence on the theoretical account of
the electron correlations -- 
{\it i.e., we want to test the accuracy of atomic calculations} --  
then this difference should be as large as possible,
$F_{ns}^{\Sigma}\gg F_{n's}^{\Sigma}$. The most suitable state to
test is therefore the ground state, with the other state chosen to be
as high as possible (see Fig.~\ref{fig:corr}). 
On the other hand, for the case where 
$F_{ns}^{\Sigma} - F_{n's}^{\Sigma}\approx 0$ (e.g., both states with high
principal quantum number), the dependence on the correlations is largely 
removed in the ratio, and we may perform 
high-precision predictions of the hyperfine structure. 
Note that it may happen that sub-1\% many-body contributions 
do not appear with the same trend across $n$ --  including, e.g., when 
fitting procedures are used and states are treated differently 
-- and the hyperfine structure for 
lower states may be evaluated more accurately than for higher states.

\section{Application of the ratio method}

We apply the ratio method to the hyperfine structure results for
states $6s$ and $7s$ of $^{133}$Cs used to test the accuracy
of the most precise calculations for atomic parity violation performed
in Refs.~\cite{dfg2002,pbd2009}. 
We are limited to these lowest $ns$ states, as only data for the
states involved in the parity violation transition were presented in these works. 
The data are shown in Table~\ref{tab:ratio}. 
We include two sets of results from
Ref.~\cite{dfg2002} corresponding to {\it ab initio} values and values
obtained with modified wave functions found by fitting the calculated binding
energies to measured data.
The original theory results are presented in the first rows. 
The results of both works 
\cite{dfg2002,pbd2009} were obtained by modeling the nuclear
magnetization distribution as a uniform sphere. 
The QED radiative corrections were not included in Ref.~\cite{dfg2002}, while
subsequently-determined, rigorous calculations for $6s$ \cite{SC_hfs} 
were included in Ref.~\cite{pbd2009} 
(with relative QED corrections for $7s$ assumed to be
equal to those for $6s$).
The percentage deviations of these values from experiment are shown in
the following columns. For the hyperfine structure for
$6s$, the deviations of the unfitted and
fitted results of Ref.~\cite{dfg2002} are $0.73\%$ and
$0.09\%$, while the deviation of the result of Ref.~\cite{pbd2009}
is $0.37\%$; for $7s$, the deviations are $-0.09\%$, $-0.37\%$,
$-0.23\%$, respectively.  
We expect that a better indication of the quality of the wave
functions may be obtained by employing the more well-motivated single-particle model for
the magnetization distribution, as well as accounting for the QED corrections 
where these have been omitted. For the $s$-states of $^{133}$Cs, adjusting the magnetization model
accordingly amounts to a correction of $0.53\%$ and inclusion of QED 
radiative effects amounts to a correction of $-0.38\%$ \cite{GVF_QED_BW}. 
With these corrections, we obtain values for the magnetic hyperfine
constants -- shown in Table~\ref{tab:ratio} -- with the following deviations from experiment 
for the unfitted and fitted results of Ref.~\cite{dfg2002}
and the results of Ref.~\cite{pbd2009}, respectively: $0.88\%$, $0.24\%$, and $0.94\%$ for
the $6s$ state; 
$0.06\%$, $-0.22\%$, and $0.34\%$ for the $7s$ state.
The corrections to the hyperfine results of Ref.~\cite{pbd2009} are 
sizeable enough to affect the error assignment for the parity
violating amplitude computed in that work.
 
The nuclear and QED effects and their uncertainties are 
difficult to quantify, however, and application of the ratio method ensures that any deviations are
associated with the uncertainties of the electronic calculations
alone.
The results for the
magnetic hyperfine constants for the $6s$ state  
of $^{133}$Cs, obtained from the ratio method with data from $7s$, are
shown in the final rows of Table~\ref{tab:ratio}; for illustration, 
we also present values for $7s$ found from the ratio involving $6s$.  
We obtain values for $6s$ with deviations $0.82\%$, $0.46\%$, and
$0.59\%$ for the unfitted and fitted data from Ref.~\cite{dfg2002} and
the data from Ref.~\cite{pbd2009}, respectively. 
In the general case, these deviations are attributable to the errors associated with {\it both}
states in the ratio. 
Note that the procedure for testing the atomic theory is equivalent to
comparing the theory ratio $A_{n\kappa}/A_{n'\kappa}$ or $A^{\rm MB}_{n\kappa}/A^{\rm MB}_{n'\kappa}$ 
with the empirical one $A^{\rm exp}_{n\kappa}/A^{\rm exp}_{n'\kappa}$. 
Expressing the many-body value $A^{\rm MB}_{n\kappa}$ in
terms of an ``exact'' value $A^{\rm MB,\,exact}_{n\kappa}$ and a
deviation from this value $\delta ^{\rm MB}_{n\kappa}=A^{\rm
  MB}_{n\kappa} -A^{\rm MB,\,exact}_{n\kappa}$, the ratio
$A_{n\kappa}^{\rm MB}/A_{n'\kappa}^{\rm MB}$ deviates from the exact
value by $(\delta A^{\rm MB}_{n\kappa}/A^{\rm
  MB}_{n\kappa}-\delta A^{\rm MB}_{n'\kappa}/A^{\rm MB}_{n'\kappa})$, 
that is, by the {\it difference} in the relative deviations
from the exact or experimental value. 
We emphasise that the dependence on QED corrections 
is removed in the ratio when both states are treated in the same way. 
Similarly, the dependence on the chosen magnetization distribution, and  
the nuclear magnetic moment, is also removed.
The relatively large deviations seen above -- {\it connected only with the 
uncertainties in the electronic wave functions of the two states} --
indicate that there is room for improvement of the wave functions and
calculations of atomic parity violation.  
It would be more useful, though, to isolate the uncertainty connected with  
{\it just one state}, that is, e.g., $(\delta A^{\rm MB}_{6s}/A^{\rm
  MB}_{6s})$ and $(\delta A^{\rm MB}_{7s}/A^{\rm MB}_{7s})$ and not
their difference. In order to do this, high precision measurements 
and calculations for high states are required, as discussed earlier in the paper. 
We propose such a prescription -- a detailed study of the hyperfine 
structure for a number of states across principal quantum number
-- in future analyses of atomic parity violation calculations.

We would also like to demonstrate the predictive power of the ratio
method. We may use the theory results of Ref.~\cite{gomez} for states $8s$ 
and $9s$ of $^{210}$Fr where the deviations from experiment 
for both states are 0.4\% and cancel in the ratio. 
These calculations were performed using a 
magnetic moment with 2\% uncertainty, 
the nuclear magnetization taken to have a Fermi distribution, and 
QED corrections omitted, with raw values $1584$\,MHz and
$624.8$\,MHz. Using the data for $8s$ (with the measurement from Ref.~\cite{simsarian}) to make a
``prediction'' for $9s$, we obtain a value for the hyperfine constant 
$622.4$\,MHz compared to the measured value $622.25(36)$\,MHz \cite{gomez}.   

\section{Conclusions}

In summary, the ratio method advocated in the current work allows the 
electronic wave functions in the nuclear
region to be probed, through hyperfine comparisons, 
free from uncertainties connected with the nuclear magnetic moment, 
nuclear magnetization distribution, and QED. 
We have demonstrated that by using this method, the remaining
electronic uncertainties are associated with the difference in the relative
errors associated with the two states. This may be reduced to the 
electronic uncertainty associated with one state by making that of the 
other very small, which we expect is possible for the highly-excited states. 
This development is important 
for the interpretation of atomic parity violation measurements in
several ways. 
Most directly, it will enable a more reliable error analysis of the hyperfine structure
to be performed, leading to a more reliable error assignment for the
theoretical parity violating amplitude and deduced nuclear weak charge. 
It may also enable improved control of the electronic wave functions
in the nuclear region and higher accuracy in calculations of atomic parity violation 
and other short-distance effects. This may be achieved 
through empirical fitting of the hyperfine structure by introducing
parameters to the correlation potential in a similar way
to the method used currently for energies. 
Far more broadly, the proposed method opens up the possibility of 
using the hyperfine structure as a benchmark against which 
developments in precision atomic many-body theory may be tested. 
Indeed, the hyperfine structure has a different sensitivity to
correlations compared to energies.

Therefore, high-precision measurements -- and calculations -- of hyperfine structure constants for
high $s$, $p_{1/2}$, and $p_{3/2}$ states of alkali-metal atoms and
alkali-metal-like ions would prove invaluable for replacing the
unknown nuclear physics parameters and QED in atomic calculations of the
hyperfine structure by means of the proposed ratio method.

\acknowledgments

We are grateful for useful discussions with V. Dzuba, A. Doherty, and
S. Bartlett and, post-submission, for stimulating discussions with L. Orozco,
J. Behr, and other members of the FrPNC Collaboration and with 
D. Elliott, J. Choi, G. Toh, and 
other members of the Purdue University group working on Cs atomic
parity violation.
J.S.M.G. would like to thank the Mainz Institute for Theoretical Physics
(MITP) for hospitality and support over a two-week visit during the 
Low-Energy Probes of New Physics program. This work was supported by
the Australian Government through the Australian Research Council
Centre of Excellence for Engineered Quantum Systems (EQuS), project number CE110001013, and 
through an Australian Research Council Future Fellowship, project number FT170100452.

\end{document}